\documentclass{ws-procs9x6-cpt16}
\begin{document}

\newcommand{\refeq}[1]{(\ref{#1})}
\def\etal {{\it et al.}}

\title{Signals for Lorentz and CPT Violation in\\
Atomic Spectroscopy Experiments and Other Systems
}

\author{Arnaldo J.\ Vargas}

\address{Physics Department, Indiana University,
Bloomington, IN 47405, USA}

\begin{abstract}
The prospects of studying nonminimal operators for Lorentz violation using spectroscopy experiments with light atoms and muon spin-precession experiments are presented.  
Possible improvements on bounds on minimal and nonminimal operators for Lorentz violation are discussed.

\end{abstract}

\bodymatter

\section{Motivation and introduction}

The Standard-Model Extension (SME) has facilitated a worldwide systematic search for Lorentz violation.  
Promptly after the introduction of the SME\cite{SME} models for Lorentz violation in spectroscopy experiments with light atoms and muon spin-precession experiments were introduced.\cite{Prev}
These models triggered experimental searches for Lorentz violation with hydrogen masers,\cite{PhHu01} muonium spectroscopy,\cite{HuPe01} and  muon spin-precession experiments.\cite{Be08}  
 
 Recently the effective Lorentz-violating hamiltonians used to obtain the models for the systems aforementioned were extended to include contributions from Lorentz-violating operators of arbitrary mass dimensions.\cite{KoMe13} 
These new hamiltonians motivated two publications.  The first publication discusses the changes, due to the introduction of nonminimal terms,  to the well-known phenomenology for Lorentz violation in spectroscopy experiments with muonic atoms and muon\footnote{
See Ref.\ \refcite{DiKo16} for a similar work with first-generation particles instead of muons.} spin-precession experiments.\cite{GoKo14} 
The other publication concentrates on the instance of spectroscopy experiments with light atoms that are composed of first-generation 
particles.\cite{KoVa15}
Note that most of the results presented in Ref.\ \refcite{KoVa15} can readily be applied to any two-fermion atom, including muonium and muonic hydrogen,
and some of the discussion here is based on this fact.

\section{Sidereal variations
}
A signal for Lorentz violation is a sidereal variation of frequencies measured in a laboratory on the surface of the Earth.  In this section we will limit attention only to effects produced by a breaking of the rotation symmetry of the laboratory frame as observed in the Sun-centered frame.  

\subsection{Bounds from previous studies}

Experimental studies that are sensitive to minimal operators might also be sensitive to nonminimal operators. 
Using the results from sidereal variation studies in muonium spectroscopy\cite{HuPe01} and muon spin-precession experiments,\cite{Be08} bounds on muon nonminimal coefficients for Lorentz violation were reported.\cite{GoKo14} 
In the future some of these bounds might be improved\cite{GoKo14} by the planned new measurements of the hyperfine structure of muonium at J-PARC, and of the antimuon anomalous frequency at J-PARC and Fermilab.
Bounds on proton and electron nonminimal coefficients for Lorentz violation were obtained\cite{KoVa15} from the results of sidereal variation studies with hydrogen masers.\cite{PhHu01} 

\subsection{Prospects and new signals}

Not all the harmonics of the sidereal frequency can contribute to the sidereal variation of the energy level of an atom. 
For two-fermion atoms such as hydrogen the maximum harmonic of the sidereal frequency that can contribute to the variation of an energy level is given by the expression $2K-1$, where $K$ is the maximum the total angular momentum $J$ of the lighter fermion and the total angular momentum $F$ of the atom.

The minimal operators for Lorentz violation can only produce variations up to the second harmonic of the sidereal frequency and contributions to variations of the energy levels with higher harmonics of the sidereal frequency are strictly due to the presence of nonminimal operators.
To study systematically the nonminimal operators, it is necessary to be sensitive to these higher harmonics by performing sidereal variation studies of transitions involving energy levels with $J>3/2$ or $F>1$. 
This suggests that some of the most promising experimental studies sensitive to  these nonminimal terms in spectroscopy experiments with light atoms, including muonium and muonic hydrogen, are sidereal variation studies of two-photon transitions such as the $2S$-$nD$ and $2S$-$nP$ transitions.\cite{KoVa15} 

The sensitive of a spectroscopy experiment to some of the coefficients for Lorentz violation can depend on the particular light atom used in the experiment.
For example, the contributions due to the nonminimal operators depend on the momentum of the fermion relative to the zero-momentum frame of the atom. 
Realizing the same experimental studies with systems with higher momentum such as deuterium and muonic hydrogen can substantially improve bounds on some of the nonminimal coefficients for Lorentz violation.\cite{GoKo14,KoVa15} 

The nonminimal terms allow contributions to the sidereal variation of the muon or antimuon anomalous frequency from all possible harmonics of the sidereal frequency.\cite{GoKo14}
The harmonics of the sidereal frequency that can contribute to the sidereal variation can be limited by restraining the mass dimensions of the operators that contribute to the energy shift to be equal to or smaller than $d$.
The maximum harmonic that can contribute in that case will be obtained by $d-2$ for even values of $d$ and $d-3$ for odd values of $d$.

The new measurement of the antimuon anomalous frequency  at Fermilab will use more energetic antimuons compared to the experiment at J-PARC\cite{prop} and for that reason it will be more  sensitive to nonminimal Lorentz-violating operators.\cite{GoKo14} This implies that the signals for Lorentz violation that the Fermilab experiment could target would include variations with harmonics higher than the second harmonic of the sidereal frequency. The J-PARC experiments would be more sensitive to the minimal coefficients for Lorentz violation\cite{GoKo14} and the targeted signals would be sidereal variations with the first and second harmonic of the sidereal frequency.

\section{Boost corrections}

Frequencies measured in a laboratory on the surface of the Earth can exhibit annual and sidereal variations due to the change of the velocity of the laboratory frame relative to the Sun-centered frame. Some of the signals for Lorentz violation presented in this section might overlap with signals presented in the previous section, however they are produced by a different set of coefficients for Lorentz violation.

The corrections due to the velocity of the laboratory frame to the 1S-2S transition and the hyperfine splitting of the ground state in hydrogen and deuterium were obtained including contributions from coefficients for Lorentz violation up to mass dimension eight.\cite{KoVa15,Ma13}  The results obtained for hydrogen can be adapted for other light atoms such as muonium, positronium, and muonic hydrogen. The signals for Lorentz violation in this case are annual variation and sidereal variation of the transition frequencies.\cite{KoVa15}

Corrections to the antimuon anomalous frequency due to the motion of the laboratory were obtained including contributions from coefficients for Lorentz violation up to mass dimension four.\cite{GoKo14} The signals are annual variation and variations up to the second harmonic of the sidereal frequency.\cite{GoKo14}

\section{Antihydrogen}

All the signals for Lorentz violation that can be studied in hydrogen can also be studied in antihydrogen.\cite{KoVa15} The planned measurements of the hyperfine splitting of the ground state of antihydrogen by the ASACUSA collaboration and the 1S-2S transitions by ALPHA and ATRAP collaborations could in the future be among the most sensitive tests discriminating between the CPT even and  the CPT odd coefficients for Lorentz violation in spectroscopy experiments. 
The study of these transitions in antihydrogen should only be the beginning.  These transitions are insensitive to CPT-violating operators that could only be studied by using transitions that involve energy levels with higher values of $F$ and $J$.  

\section*{Acknowledgments}
This work was supported 
by Department of Energy grant {DE}-SC0010120
and by the Indiana University Center for Spacetime Symmetries.

\end{document}